\newcommand{\diracslash}[1]{#1\llap{/\kern2pt}}

\newcommand{\be}{\begin{equation}}
\newcommand{\ee}{\end{equation}}
\newcommand{\bea}{\begin{eqnarray}}
\newcommand{\eea}{\end{eqnarray}}
\newcommand{\ba}[1]{\begin{array}{#1}}
\newcommand{\ea}{\end{array}}

\newcommand{\bt}{\begin{tabular}}
\newcommand{\et}{\end{tabular}}

\newcommand{\beas}{\begin{eqnarray*}}
\newcommand{\eeas}{\end{eqnarray*}}

\documentclass[preprint,prd,aps,floats,nofootinbib,floatfix]{revtex4}
\usepackage{graphicx}
\begin{document}

\title{Charmonia decay widths in magnetized matter
using a model for composite hadrons}
\author{Amruta Mishra}
\email{amruta@physics.iitd.ac.in}
\affiliation{Department of Physics, Indian Institute of Technology, Delhi,
Hauz Khas, New Delhi -- 110 016, India}

\author{S.P. Misra}
\email{misrasibaprasad@gmail.com}
\affiliation{Institute of Physics, Bhubaneswar -- 751005, India} 

\begin{abstract}
The decay widths of the charmonium states to
$D\bar D$ in isospin asymmetric nuclear matter in the presence 
of a magnetic field are studied, using a field theoretical 
model for composite hadrons with quark/antiquark constituents.
The medium modifications of these partial decay widths
arise due to the changes in the masses of the decaying 
charmonium state and the produced $D$
and $\bar D$ mesons in the magnetized hadronic matter,
calculated within a chiral effective model.
The decay widths are computed using the light 
quark--antiquark pair creation term of the free 
Dirac Hamiltonian in terms of the
constituent quark field operators.   
The results of the present investigation
are compared with the in-medium decay widths obtained 
within the $^3P_0$ model. Within the $^3P_0$ model,
the charmonium decay widths are calculated 
using the creation of a light quark--antiquark pair
in the $^3P_0$ state. 
In the presence of a magnetic field, the Landau level 
contributions give rise to positive shifts
in the masses of the charged $D$ and $\bar D$ mesons.
This leads to the decay of charmonium to the charged 
$D^+ D^-$ to be suppressed as compared to the
neutral $D\bar D$ pair in symmetric nuclear matter,
whereas in asymmetric nuclear matter, the larger mass
drop of the $D^+D^-$ pair, as compared to the $D^0 \bar {D^0}$ pair 
leads to the production of charged open charm meson pairs
to be enhanced as compared to the charmonium decay 
channel to $D^0 {\bar {D^0}}$.
\end{abstract}

\maketitle

\def\bfm#1{\mbox{\boldmath $#1$}}
\def\bfs#1{\mbox{\bf #1}}

\section{Introduction}
The properties of hadrons at high temperatures and/or 
densities comprise an important area of research,
due to their relevance in the context of the ongoing and 
future ultra relativistic heavy ion collision 
experiments at various high energy particle accelerators,
as well as in the study of the bulk matter of
astrophysical objects, e.g., neutron stars.
The estimated huge magnetic fields at Relativistic Heavy
Ion Collider (RHIC) at BNL and 
Large Hadron Collider (LHC) at CERN
\cite{Tuchin_Review_Adv_HEP_2013},
have recently initiated a lot of work on the study
of the hadrons in the presence of strong magnetic fields.
The time evolution of the magnetic field
\cite{Tuchin_Review_Adv_HEP_2013},
however, needs the solutions of the magnetohydrodynamic equations,
with a proper estimate of the electrical conductivity of the medium,
and is still an open question. Recently, there has been
a lot of work on the heavy flavour hadrons 
\cite{Hosaka_Prog_Part_Nucl_Phys}, 
and, due to the attractive interaction of the $J/\psi$
in nuclear matter \cite{leeko,amarvdmesonTprc,amarvepja},
the possibility of the $J/\psi$ forming bound states
with nuclei have been predicted 
\cite{QMC_Krein_Thomas_Tsushima_Prog_Part_Nucl_Phys_2018}.
The heavy quarkonium (charmonium and bottomonium) states
have been studied using the potential models 
\cite{eichten_1,eichten_2,Klumberg_Satz_Charmonium_prod_review,Quarkonia_QGP_Mocsy_IJMPA28_2013_review,repko}.
The heavy quarkonium state, as a bound state of the heavy quark, $Q$
and heavy antiquark ($\bar Q$) interacting by color Coulomb potential, 
has been studied in the presence of gluonic field
\cite{pes1,pes2,voloshin}. Assuming the separation bewtween 
$Q$ and $\bar Q$ to be small compared to the scale of gluonic fluctuations,
in the leading order, the mass of the quarkonium state 
is observed to be proportional to the gluon condensate 
\cite{pes1,pes2,voloshin}.
The heavy flavour mesons have been studied in the literature
using the quark meson coupling (QMC) model, 
the QCD sum rule approach 
the coupled channel approach, 
a chiral effective model,
using heavy quark symmetry and interaction
of these mesons with nucleons via pion exchange,
using heavy meson effective theory
\cite{Hosaka_Prog_Part_Nucl_Phys}. 
The open charm mesons 
\cite{Gubler_D_mag_QSR,machado_1,B_mag_QSR,dmeson_mag}
as well as the charmonium states
\cite{charmonium_mag_QSR,charmonium_mag_lee,charmonium_mag} 
have also been studied in the presence of strong magnetic fields. 
The mixing of the pesudoscalar charmonium states with the
longitudinal components of the vector charmonium states
in the presence of strong magnetic fields,
is observed to increase (decrease) 
the masses of the vector (pesudoscalar) 
charmonium states \cite{Suzuki_Lee_2017,Alford_Strickland_2013}.
In Ref. \cite {Suzuki_Lee_2017}, a study of the effects of the spin mixing
on the formation times of the charmonium states, is
observed to lead to delayed (earlier) 
formation times for the vector (psuedoscalar) charmonium states.
If the formation of $J/\psi$ is delayed, then the heavy quarks might
pass through the medium before thermalization and the $J/\psi$,
being unaffected by the thermal medium, can have a higher survival 
probability \cite{Suzuki_Lee_2017}.

The chiral SU(3) model based on a non-linear realization
of chiral symmetry and broken scale invariance \cite{paper3}, 
generalized to SU(4) to include the charm sector,
is used to study the masses of the charmonium and open charm mesons
in a hadronic medium in the presence of a magnetic field.
The charmonium masses are calculated from
the medium modification of a scalar dilaton field 
which mimics the gluon condensates of QCD 
\cite{amarvdmesonTprc,amarvepja,charmonium_mag}
and the in-medium masses of the open charm 
($D$, $\bar D$) mesons, are computed from their 
interactions with the baryons and scalar mesons 
\cite{amarvdmesonTprc,amarvepja,dmeson_mag}
in the (magnetized) hadronic medium.
The in-medium decay widths of the charmonium states to 
$D\bar D$ at zero magnetic field,
have been studied using a model with light quark pair 
created in the $^3P_0$ state \cite{amarvepja,3p0,3p0_1,friman}, 
as well as, using a field theoretical model with composite hadrons
\cite{amspmwg}.
The results of the present investigation of the charmonium
decay widths in magnetized nuclear matter calculated using 
the model for composite hadrons are compared 
with the results obtained using the $^3P_0$ model 
\cite{charm_decay_mag_3p0}.

The outline of the paper is as follows : In section II, we describe
briefly the field theoretical model with composite hadrons
with quark/antiquark constituents, used in the present 
work to compute the partial decay widths of the charmonium
states to $D\bar D$ in magnetized hadronic matter.
These in-medium decay widths are computed from the
mass modifications of the decaying charmonium state
and the produced $D$ and $\bar D$ mesons, calculated
within a chiral effective model.
In section III, we discuss the results obtained in the present
investigation of these in-medium charmonium  decay widths.
In section IV, we summarize the findings of the present study.
The salient features of the field theoretic model for composite
hadrons are presented in Appendix A, and the chiral effective model
for study of the charmed mesons is breifly descirbed in 
Appendix B.

\section{Decay width of Charmonium state to $D\bar D$ 
within a model for composite hadrons}

We use a field theoretical model 
for composite hadrons with quark/antiquark constituents
\cite{spm781,spm782,spmdiffscat} to study 
the charmonium decay widths to $D\bar D$
in isospin asymmetric nuclear matter in the
presence of a magnetic field.
The model has been used to investigate these 
charmonium decay widths in hadronic matter
in the absence of a magnetic field \cite{amspmwg}.
With explicit constructions of the charmonium 
state and the open charm mesons, the decay width is
calculated using the quark antiquark pair creation
term of the free Dirac Hamiltonian 
for constituent quark field \cite{amspmwg}.
The model for the  composite hadrons 
with quark constitutents has been described in Appendix A.

The relevant part of the quark pair creation term is through the
$d \bar d (u \bar u)$ creation for decay of the charmonium
state, $\Psi$, to the final state $D^+D^-$($D^0 {\bar {D^0}}$). 

For $\Psi \rightarrow D^+({\bf p}) D^- ({\bf p'})$,
this pair creation term is given as 
\begin{equation}
{\cal H}_{d^\dagger\tilde d}({\bf x},t=0)
=Q_{d}^{(p)}(\vec x)^\dagger (-i{\bf {\alpha}}\cdot
{\bf \bigtriangledown} +\beta M_d)
{\tilde Q}_d^{(p')}(\vec x) 
\label{hint}
\end{equation}
where, $M_d$ is the constituent mass of the $d$ quark.
The subscript $d$ of the field operators in equation (\ref{hint})  
refers to the fact that the $\bar d$ and $d$ are the constituents 
of the $D^+$ and $D^-$ mesons with momenta $\vec p$ and
$\vec p'$ respectively in the final state of the
decay of the charmonium state, $\Psi$.

The charmonium state, $\Psi$ ($J/\psi$,$\psi'\equiv\psi(3686)$,
$\psi^{''}\equiv\psi(3770))$ 
with spin projection m, at rest is written as
\begin{equation}
|\Psi_m({\bf 0})\rangle = {\int {d {\bf k} {c_r} ^i ({\bf k})^\dagger
u_r a_m(\Psi,{\bf k})\tilde {c_s}^i (-{\bf k})v_s|vac\rangle}},
\label{Psi}
\end{equation}
where, $i$ is the color index of the quark/antiquark operators.
The wave functions for the charmonium states, $J/\psi$,
$\psi (3686)$ and $\psi (3770)$ are assumed to be
harmonic oscillator wave functions,
corresponding to the 1S, 2S and 1D states
respectively. With this assumption, 
the expressions for $a_m (\Psi,\bf k)$
are given as \cite{spmddbar80}
\begin{equation}
a_m(\Psi,\bfs k)
=\bfm\sigma_m\frac{1}{\sqrt{6}}
\left (\frac {{R_{\Psi}}^2}{\pi} \right)^{3/4}
\exp \Big(-\frac {{R_{\Psi}}^2 {{\bf k}}^2}{2}\Big),
\label{amjpsi}
\end{equation}
for $\Psi\equiv J/\psi$,
\begin{equation}
a_m(\Psi,\bfs k)
=\bfm\sigma_m\frac{1}{2}
\Big({\frac{R_{\Psi}^2}{\pi}}\Big)^{3/4}
\left(\frac{2}{3}R_{\Psi}^2\bfs k^2-1\right)
\exp\Big(-\frac{R_{\Psi}^2 {\bfs k}^2}{2}\Big).
\label{ampsip}
\end{equation}
for $\Psi \equiv \psi(3686)$,
and, 
\begin{eqnarray}
a_m(\Psi,{\bf k})&=&\frac {1}{3\sqrt {5\pi}} 
\pi^{-{1}/{4}}
(R_{\Psi}^2)^{7/4} {{\bf k}}^2 
\Big ( \bfm\sigma_m -3 (\bfm\sigma \cdot \hat {k})\hat {k}^m \Big )
\nonumber \\
&\times &
\exp \Big (-\frac{1}{2} {R_{\Psi}}^2 
{{\bf k}}^2\Big),
\label{ampsipp}
\end{eqnarray}
for $\Psi \equiv \psi (3770)$.
In equations (\ref{amjpsi}), (\ref{ampsip}) and (\ref{ampsipp}),
$R_\Psi$ corresponds to the strength of the harmonic oscillator 
potential for the charmonium state, $\Psi$.
The parameters $R_{J/\psi}$, $R_{\psi'}$, 
$R_{\psi''}$, are fitted to their rms radii to be
(0.47 fm)$^2$, (0.96 fm)$^2$ and 1 fm$^2$ respectively, 
which yield their values as (520 MeV)$^{-1}$, (390 MeV)$^{-1}$
and (370 MeV)$^{-1}$ \cite{leeko}. 

The $D(D^+,D^0)$ and $\bar D(D^-,\bar {D^0})$ states, 
with finite momenta are contructed in terms of the
constituent quark field operators obtained 
from the quark field operators of these mesons 
at rest through a Lorentz boosting \cite{spmdiffscat} 
(see Appendix A). These states are explicitly 
given as
\begin{eqnarray}
|D ({\bf p})\rangle  = \int u_{D}({\bf k}_2) 
{c_r}^{{i_1}}({\bf k}_2+\lambda_2 {\bf p})
^\dagger u_r^\dagger 
\tilde {q_s}^{i_1} 
(-{\bf k}_2 +\lambda_1 {\bf p})v_s
d\bfs k_2
\label{d}
\end{eqnarray}
and 
\begin{eqnarray}
|{\bar D} ({\bf p}')\rangle 
=  \int u_{{\bar D}}({\bf k}_3) 
{q_r}^{{i_2}}({\bf k}_3+\lambda_1 {\bf p}')
^\dagger u_r^\dagger 
\tilde {c_s}^{i_2} 
(-{\bf k}_3 +\lambda_2 {\bf p}') v_s
d\bfs k_3.
\label{dbar}
\end{eqnarray}
In the above,  $q=(d,u)$ for $(D^+,D^-)$ and  $(D^0,\bar {D^0})$
respectively, and, we assume harmonic oscillator wave functions 
for the $D$ and $\bar D$ mesons given as
\begin{equation}
u_{D (\bar D)}({\bf k})=\frac{1}{\sqrt{6}}
\Big (\frac {{R_D}^2}{\pi} \Big)^{3/4}
\exp\Big(-\frac {{R_D}^2 {{\bf k}}^2}{2}\Big),
\label{uddbar}
\end{equation}
where the harmonic oscillator strength  
parameter for $D(\bar D)$ meson is taken
as $R_D$=(310 MeV)$^{-1}$ so as to give the experimental
values of the vacuum decay widths of 
$\psi (3770)\rightarrow D\bar D$,
and, $\psi(4040)$ to $D\bar D$, $D\bar D^*$, $D^* \bar D$ and 
$D^*\bar D^*$ \cite{leeko}. 

In equations (\ref{d}) and (\ref{dbar}), 
$\lambda_1$ and $\lambda_2$ are the 
fractions of the mass (energy) of the $D(\bar D)$ meson
at rest (in motion), carried by the constituent 
light (d,u) antiquark (quark)
and the constituent heavy charm quark (antiquark), 
with $\lambda_1 +\lambda_2=1$.
These are calculated by assuming the binding energy of the hadron 
as shared by the quark(antquark) are inversely 
proportional to the quark (antiquark) masses
\cite{spm782}. The energies of the
the light antiquark (quark) and heavy charm quark (antiquark), 
$\omega_i=\lambda_i m_D (i=1,2)$, are assumed to be 
\cite{amspmwg,spm782}
\begin{eqnarray}
\omega_1=M_q+ \frac{\mu}{M_q}\times BE,\;\; 
\omega_2=M_c+ \frac{\mu}{M_c}\times BE,
\end{eqnarray}
where $BE=(m_D-M_c-M_q)$ is the binding energy of $D (\bar D)$
meson, with $M_c$ and $M_q$ as the masses of the
constitutent charm and light quark (antiquark), and,
$\mu$ is the reduced mass of the heavy-light quark-antiquark 
system (the $D (\bar D)$ meson), defined by $1/\mu=1/M_q+1/M_c$.
The reason for making this assumption comes from the example of
hydrogen atom, which is the bound state of the proton and the electron.
As the mass of proton is much larger as compared to the mass of the 
electron, the binding energy contribution from the electron is
$\frac{\mu}{m_e}\times BE \simeq BE$ of hydrogen atom, and the
contribution from the proton is $\frac{\mu}{m_p}\times BE$, 
which is negligible as compared to the total 
binding energy of hydrogen atom, since $m_p >> m_e$. 
With this assumption, the binding energies of the heavy-light mesons, 
e.g., $D$ and $(\bar D)$ mesons \cite{amspmwg}
and, $B$ $\bar B$ mesons \cite{amspm_upsilon}, 
mostly arise from the contribution from the light quark (antiquark).

The expression of the decay width of the 
charmonium state, $\Psi$ to $D\bar D$,
as calculated in the present model for composite hadrons
is given by equation (\ref{gammapsiddbar}).
The parameter, $\gamma_\Psi$, in the expression for 
the charmonium decay width,
is a measure of the coupling strength
for the creation of the light quark antiquark pair,
to produce the $D\bar D$ final state. 
This parameter is adjusted to reproduces the
vacuum decay widths of $\psi(3770)$ to $D^+D^-$ and
$D^0 \bar {D^0}$ \cite{amspmwg}. 
The decay width of the charmonium state is observed to have
the dependence on the magnitude of the 3-momentum
of the produced $D(\bar D)$ meson, $|{\bf p}|$, 
as a polynomial part multiplied by an exponential 
term.  
The medium modification of the charmonium decay width
is studied due to the mass modifications of the
charmonium state, the $D$ and $\bar D$ mesons through 
$|{\bf p}|$, which is given as, 
\begin{equation}
|{\bf p}|=\Big (\frac{{m_\psi}^2}{4}-\frac {{m_D}^2+{m_{\bar D}}^2}{2}
+\frac {({m_D}^2-{m_{\bar D}}^2)^2}{4 {m_\Psi}^2}\Big)^{1/2}.
\label{pd}
\end{equation}
In equation (\ref{pd}), the masses of the charmonium and
open charm mesons are the effective masses in the
hadronic matter in the presence of a magnetic field.
These in-medium masses are calculated using a chiral 
effective model, which is briefly described in Appendix B.

The chiral effective model incorporates the broken scale invariance
of QCD, as has already been mentioned, through a scalar dilaton field,
which mimics the gluon condensates of QCD \cite{paper3}. 
The charmonium masses are calculated from 
the medium modifications of the gluon condensates,
which is obtained from the medium changes of the scalar dilaton field 
\cite{amarvepja,charmonium_mag}.
The mass modifications of the $D$ and $\bar D$ mesons
are calculated from their interactions with the
nucleons and the scalar mesons in the magnetized nuclear
medium \cite{amarvepja,dmeson_mag}, within the chiral effective model.
The proton, which is the charged nucleon, has 
contributions from the Landau energy levels.
The anomalous magnetic moments of the nucleons are considered
in the present study. The charged $D^\pm$ mesons have  
positive contributions in the masses in the presence 
of a magnetic field, and we account for the lowest Landau 
level contribution to the masses of these charged 
open charm mesons \cite{dmeson_mag}. 

\section{Results and Discussions}

In the present work, the charmonium decay widths to $D\bar D$
in magnetized nuclear matter are investigated within 
a field theoretical model for composite hadrons with quark/antiquark
constituents. The medium modifications of these decay widths
are computed from the  changes in the masses of the 
charmonium state, $\Psi$, and the open charm mesons,
calculated within a chiral effective model
\cite{dmeson_mag,charmonium_mag,charm_decay_mag_3p0}.
The charmonium decay width, $\Gamma (\Psi \rightarrow D\bar D)$
is calculated using the light quark
pair creation term of the free Dirac Hamiltonian 
expressed in terms of the constituent quark
operators, using explicit constructions
for the charmonium and the open charm ($D$,$\bar D$) mesons. 
The matrix element for the calculation
of the decay width is multiplied with a factor
$\gamma_\psi$, which gives the strength of the
light quark pair creation leading to the decay
of the charmonium state to $D\bar D$ in the 
magnetized hadronic medium \cite{amspmwg}. 
The value of $\gamma_\psi$
to be 1.35, is chosen so as to reproduce the decay widths
of $\psi(3770) \rightarrow D^+ D^-$ and   
$\psi(3770) \rightarrow D^0 \bar{D^0}$ in vacuum,
to be around 12 MeV and 16 MeV respectively 
\cite{amspmwg}. The constituent quark masses
for the light quarks ($u$ and $d$) are taken to 
be 330 MeV and for the charm quark, the value
is taken to be $M_c=1600$ MeV \cite{amspmwg}.

The effects of isospin asymmetry, density
and magnetic field on the partial decay
width of $\Psi \rightarrow D\bar D$ 
are investigated within the composite model for the hadrons.
The charmonium masses are studied in the 
magnetized nuclear matter within the
chiral effective model, due to the medium
changes in the scalar dilaton field, $\chi$,
which mimics the gluon condensates of QCD 
\cite{charmonium_mag}.
The masses of the open charm mesons in the magnetized 
nuclear matter are modified due to their
interactions with the nucleons as well as
the scalar mesons, $\sigma$, $\zeta$,
and $\delta$ \cite{dmeson_mag}. 
These scalar fields and the dilaton field $\chi$ 
are solved from their coupled equations of motion. 
It might be noted here that
the $D$ and $\bar D$ mesons in Ref. \cite{dmeson_mag}
as well as the kaons and antikaons \cite{kmeson_mag}
were studied in the frozen glueball approximation,
i.e. the value of $\chi$ was fixed at its vacuum value.
This approximation was chosen as the modifications 
of the open charm (strange) mesons arise 
due to the scalar mesons, $\sigma$, $\zeta$ and $\delta$, 
which do not have large modifications, when the
medium dependence of the scalar dilaton field,
$\chi$ is taken into consideration as well.
The $D$ and $\bar D$ masses in the magnetized nuclear
matter, including the variation of the dilaton field 
has been considered in Ref. \cite{charmonium_mag}
to study the mass modifications of the charmonium states
(which are calculated from the medium changes of the 
$\chi$ field), as well as in Ref. \cite{charm_decay_mag_3p0},
where the partial decay widths of charmonium states
to $D\bar D$ have been studied using the $^3P_0$ model.
The present study also uses the medium dependence
of the dilaton field to calculate the 
charmonium decay widths to $D\bar D$ using a
field theoretical model for composite hadrons.
For given values of the baryon density, $\rho_B$,
isospin asymmetry parameter, $\eta=(\rho_n-\rho_0)/(2\rho_B)$,
and magnetic field, the values of the scalar fields are 
solved from their equations of motion, within the chiral 
effective model, which are then used to calculate the in-medium
masses of the charmonium states and the open charm mesons.

\begin{figure}
\hskip -0.4in
\includegraphics[width=12cm,height=12cm]{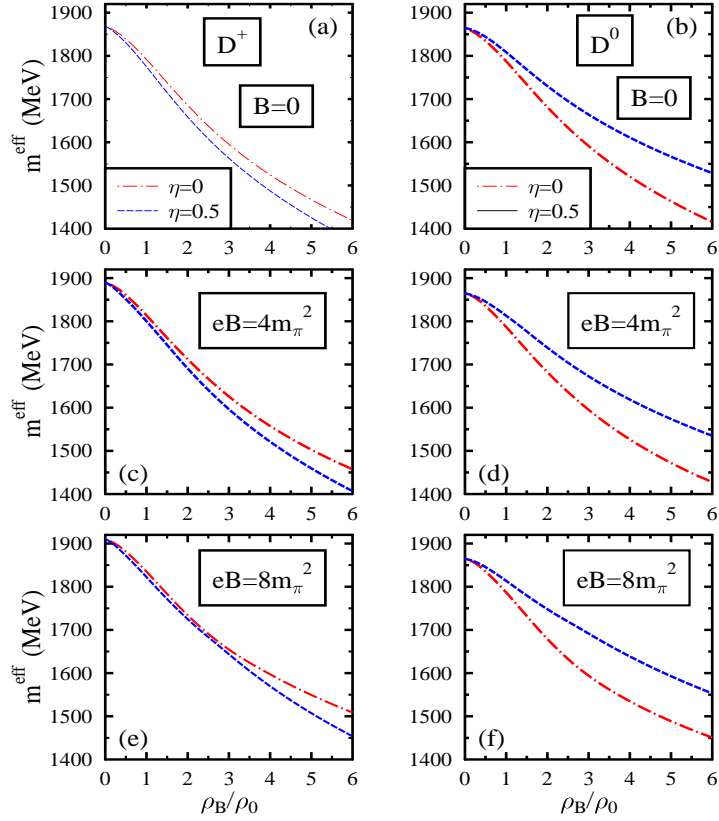}
\caption{(Color online)
In-medium masses of $D^+$ (in panels (a), (c) and (e)) 
and $D^0$ (in panels (b), (d) and (f)),
as calculated using the chiral effective model,
are plotted as functions of $\rho_B/\rho_0$,
for given values of the magnetic field for symmetric ($\eta$=0) 
and asymmetric (with $\eta$=0.5) nuclear matter.
}
\label{md_mag_eB}
\end{figure}

\begin{figure}
\hskip -0.4in
\includegraphics[width=12cm,height=12cm]{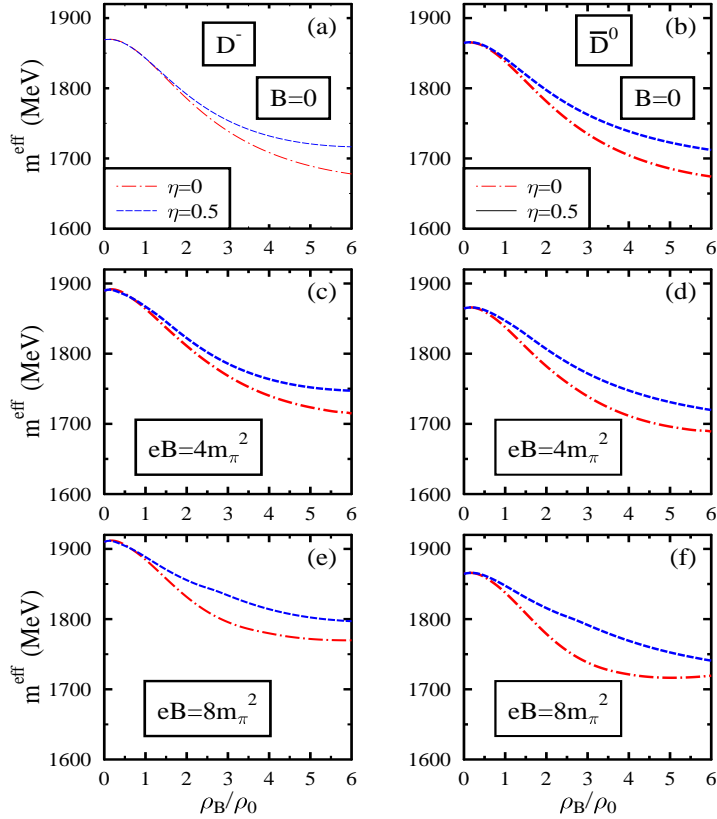}
\caption{(Color online)
Same as fig. \ref{md_mag_eB}, for $D^-$ and $\bar {D^0}$.
}
\label{mdbar_mag_eB}
\end{figure}

\begin{figure}
\hskip -0.4in
\includegraphics[width=12cm,height=12cm]{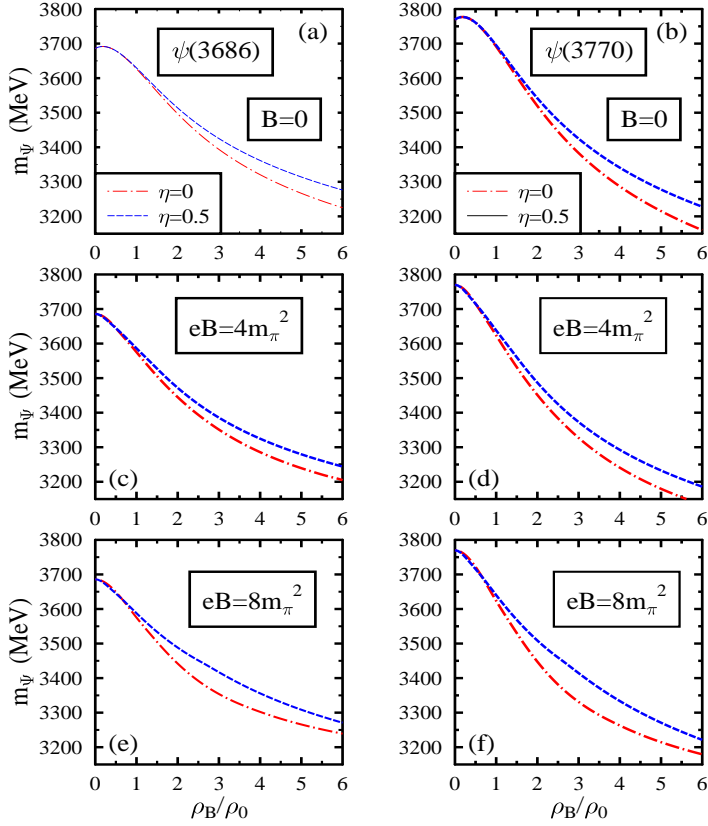}
\caption{(Color online)
Same as fig. \ref{md_mag_eB}, 
for $\psi (3686)$ and $\psi (3770)$.
}
\label{mpsi_eB_2s_1d}
\end{figure}

\begin{figure}
\hskip -0.4in
\includegraphics[width=12cm,height=12cm]{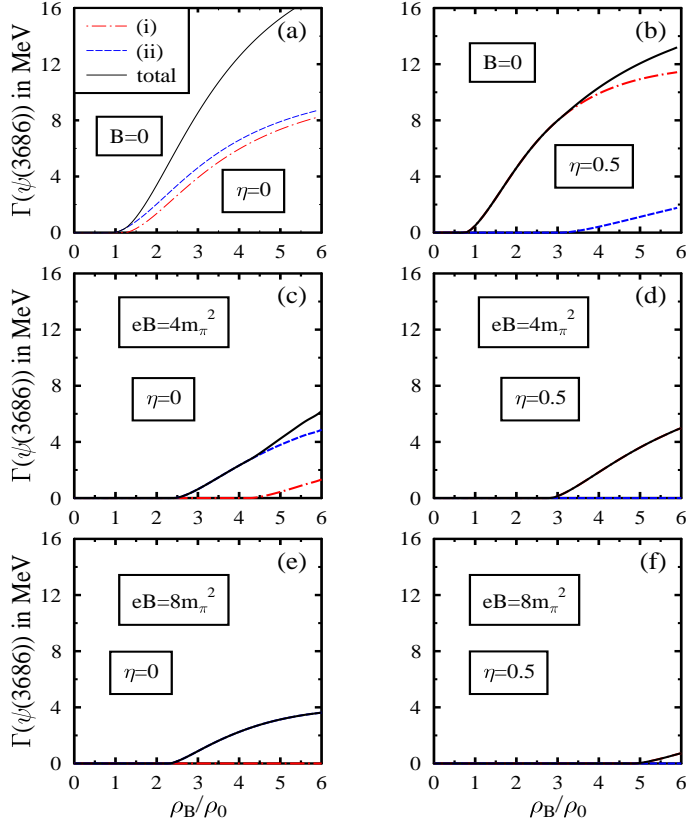}
\caption{(Color online)
Decay widths of $\psi(3686)$ to (i) $D^+D^-$, 
(ii) $D^0\bar {D^0}$, and the total of these two channels ((i)+(ii)),
as calculated using the field theoretic model for composite hadrons,
are plotted as functions of $\rho_B/\rho_0$,
for given values of the magnetic field for symmetric ($\eta$=0) 
and asymmetric (with $\eta$=0.5) nuclear matter.
}
\label{dwFT_psi3686_mag}
\end{figure}

\begin{figure}
\hskip -0.4in
\includegraphics[width=12cm,height=12cm]{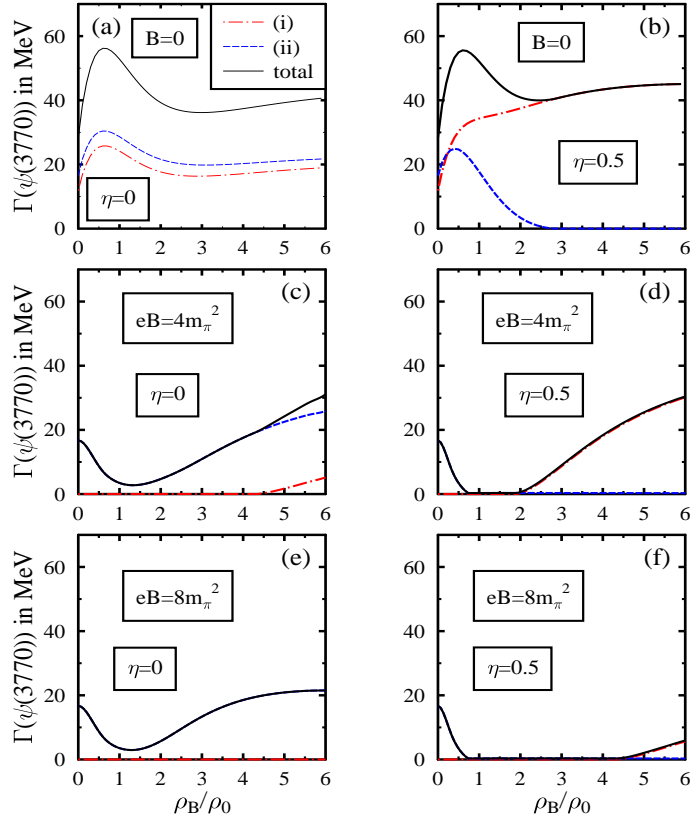}
\caption{(Color online)
Same as fig. \ref{dwFT_psi3686_mag}, 
for decay widths of $\psi(3770)$.
}
\label{dwFT_psi3770_mag}
\end{figure}

\begin{figure}
\hskip -0.4in
\includegraphics[width=12cm,height=12cm]{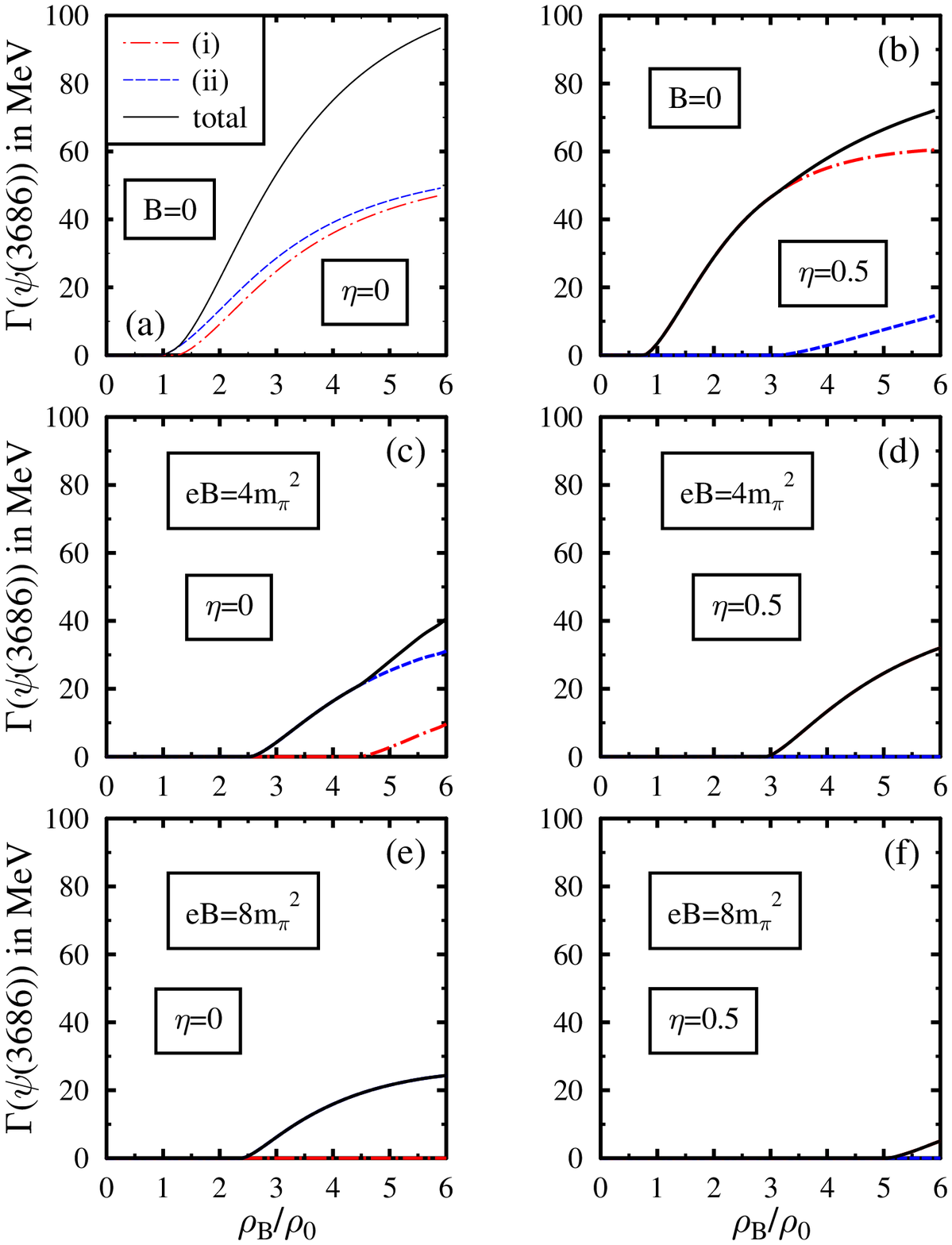}
\caption{(Color online)
Same as fig. \ref{dwFT_psi3686_mag}, within $^3P_0$ model.}
\label{psi3686_decay_3p0_mag}
\end{figure}

\begin{figure}
\hskip -0.4in
\includegraphics[width=12cm,height=12cm]{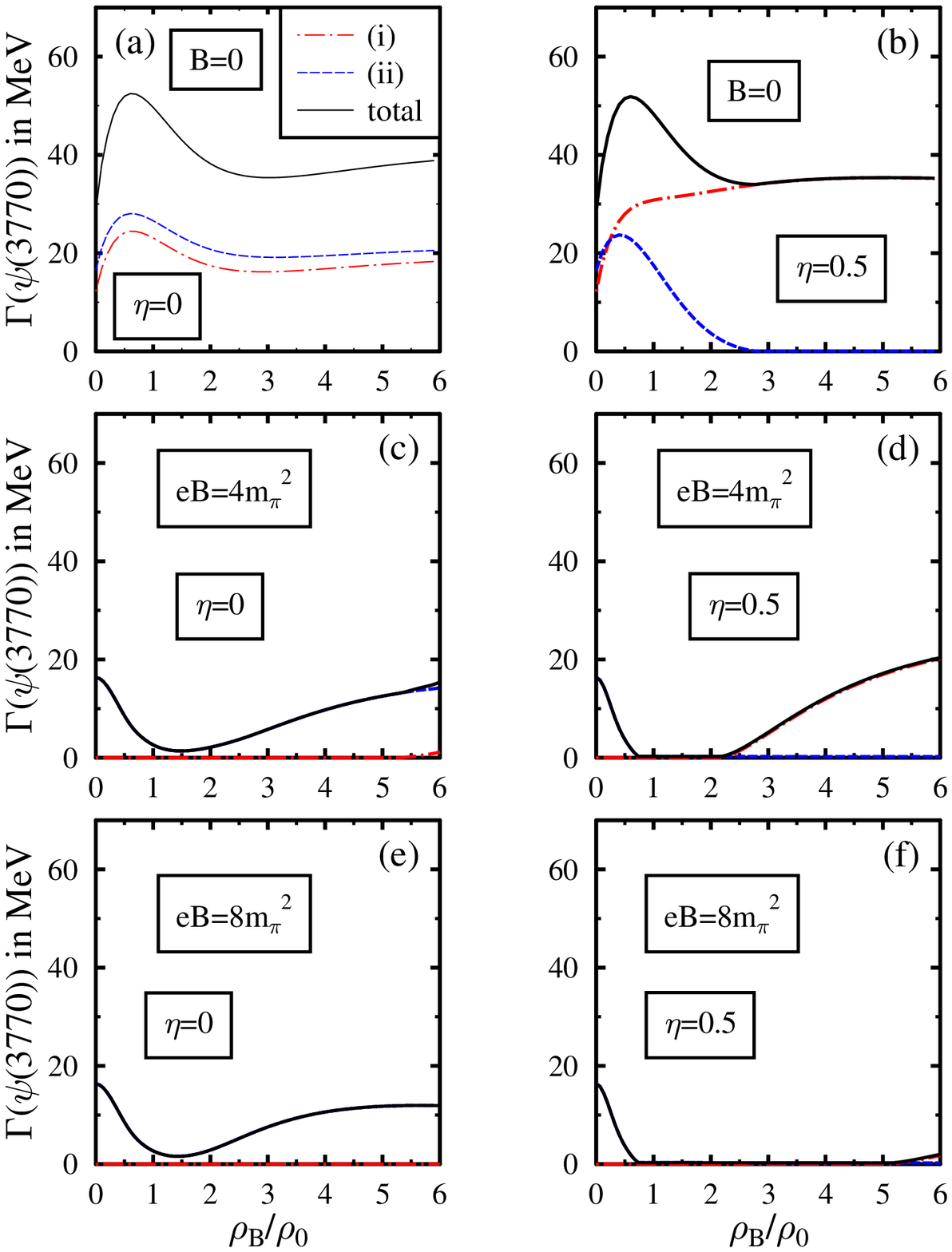}
\caption{(Color online)
Same as fig. \ref{dwFT_psi3770_mag}, within $^3P_0$ model.}
\label{psi3770_decay_3p0_mag}
\end{figure}

The charmonium decay widths are modified in the medium 
due to the mass modifications of the charmonium state
and the open charm mesons, calculated
using a chiral effective model 
\cite{dmeson_mag,charmonium_mag,charm_decay_mag_3p0} 
(see Appendix B).
In the absence of a magnetic field, the charmonium decay widths
to $D\bar D$ have been studied using the model for composite 
hadrons as used in the present work \cite{amspmwg}.
It was observed that the decay of $J/\psi$ to 
$D\bar D$ is possible, for densities higher than 
4--4.5$\rho_0$, for zero magnetic field. 
However, in the presence of a magnetic field, for $eB=4 m_\pi^2$
and $eB=8 m_\pi^2$ as considered in the present work,
the decay of $J/\psi$ to $D\bar D$ does not become kinematically
possible.  

For the sake of completeness, we show the
effects of the magnentic field, baryon density
as well as isospin asymmetry on the masses
of the $D (D^+,D^0)$, $\bar D (D^-, {\bar {D^0}})$ 
mesons and the charmonium 
states ($\psi(3686)$ and $\psi(3770)$)
in figures \ref{md_mag_eB}, \ref{mdbar_mag_eB}
and  \ref{mpsi_eB_2s_1d}, respectively.
The masses obtained in presence of a magnetic field
are compared to the case of zero magnetic field,
shown in panels, (a) and (b) in these figures,
for the symmetric and asymmetric nuclear matter.
The charmonium masses are calculated from the 
in-medium values of the scalar gluon condensate,
obtained from the medium changes of the dilaton 
field within the chiral effective model,
as has already mentioned. 
The values of $-9.3$, $-126.4$ and $-167.5$ MeV
for the mass shifts for the charmonium states,
$J/\psi$, $\psi (3686)$ and $\psi(3770)$
at $\rho_B=\rho_0$, may be compared to 
the values of $-8$, $-100$ and $-140$ MeV 
calculated using the linear density 
approximation \cite{leeko}.

The model for composite hadrons with quark/antiquark constituents,
used in the present work of calculation of
the charmonium decay widths is descirbed briefly 
in Appendix A. The expression for the charmonium decay width 
as given by equation (\ref{gammapsiddbar}), 
has dependence on the in-medium masses
of the charmonium state and the open charm mesons,
through the magnitude of the 3-momentum of the $D (\bar D)$
meson, given by equation (\ref{pd}). 
The dependence of the decay width on $|{\bf p}|$
is a polynomial part multiplied by an exponential
function in $|{\bf p}|$.

%
The $^3P_0$ model 
\cite{3p0,3p0_1,friman} describes the charmonium decay
through a light quark antiquark pair creation 
in the $^3P_0$ state. The created quark (antiquark)
combine with the charm antiquark (quark) to form the
$D$ and $\bar D$ mesons. The matrix element for
a general decay $A\rightarrow B C$ is taken as
$M_{A\rightarrow BC}
=\langle BC |\gamma ({\bar {q_s}} q_s)^{^3P_0}|A\rangle $,
where $\gamma$ is a coupling parameter, which characterizes 
the probability of creating a quark-antiquark pair.
The parameter $\gamma$ is fitted to the experimentally
observed decay width of $A\rightarrow BC$.
The decay amplitudes for a variety of decay processes
have been listed in Ref. \cite{3p0_1}.
A formulation which is equivalent to the
$^3P_0$ model is due to an interaction Hamiltonian
involving quark fields given as 
${\cal H}_I=g \int d {\bf x} \bar \psi ({\bf x}) \psi ({\bf x})$
\cite{3p0}.
The quark-antiquark pair creation term of this
interaction Hamiltonian, in the non-relativistic limit, 
gives the decay amplitudes to be same as obtained using
the $^3P_0$ model, with the identification
$\gamma=\frac{g}{2 m_q}$, where $m_q$ is the mass
of the light quark field. A typical value of $m_q$ 
is 330 MeV \cite{3p0_1},
which is also constituent quark mass
of the light quark (u,d) chosen in the present work
of charmonium decay widths using the model for the
composite haddrons with quark/antiquarks.
The $^3P_0$ model does not account for the color.
However, taking the color of the quark into account
will lead to a multiplying factor,
which would simply lead to a redefinition
of the parameter, $\gamma$ (or equivalently $g$).

The changes in the decay widths of the charmonium states 
to the $D\bar D$ in nuclear matter were studied in 
Ref. \cite{friman} using the $^3P_0$ model, 
arising due to mass drop of the $D$ meson in the medium. 
These were studied as functions of the mass of the $D$ 
(assuming the mass of the $\bar D$ meson to be identical 
with the $D$ meson mass). The decay widths of
$\psi(3686)$, $\psi(3770)$ and $\chi_{c0}$ to $D\bar D$ 
were observed to increase initially with drop in the
mass of the $D$ meson, and then decreased
with further drop in the $D$ meson mass reaching
a value of zero (so called nodes). This was followed by
an increase with further drop in the $D$ meson mass. 
This kind of behaviour for the charmonium decay widths
were observed using the $^3P_0$ model, when 
the masses of the $D$ and $\bar D$ 
mesons are calculated using the chiral effective model
\cite{amarvepja}, and the mass modifications
of the charmonium states are neglected.
However, the vanishing of the decay widths were no longer
observed when the mass modifications of the charmonium
states were also considered \cite{amarvepja}.
The behaviour of the
decay width has a form of a polynomial function
multiplied by an exponential function of $|\bf p|$,
given by equation (\ref{pd}), in terms of the masses
of the charmonium state and the open charm mesons.
The effects of magnetic fields on the charmonium decay
widths calculated using the $^3P_0$ model due to the
mass modifications of the charmonium, $D$ and $\bar D$
mesons were also studied in Ref. \cite{charm_decay_mag_3p0}.
The results of
the decay widths calculated using the model for
composite hadrons (as shown in figures 
\ref{dwFT_psi3686_mag} and \ref{dwFT_psi3770_mag}) 
as used in the present work,
are compared with the results obtained using the
the $^3P_0$ model (shown in figures \ref{psi3686_decay_3p0_mag}
and \ref{psi3770_decay_3p0_mag}). 
In both the $^3P_0$ model, as well as, in the model
for the composite hadrons as in the present work,
the medium dependence of the charmonium decay width
is through $|{\bf p}|$, and the expression for the
charmonium decay width turns out to be the form 
of a polynomial function multiplied
by an exponential part, calculated in the respective models.

The effects of magnetic field, isospin asymmetry
and density on the decay widths of 
$\psi(3686)$ to $D\bar D$ and
$\psi(3770)$ to $D\bar D$, as calculated in the present work, 
are plotted in figures \ref{dwFT_psi3686_mag}
and \ref{dwFT_psi3770_mag} respectively.
These decay widths are shown as functions 
of the baryon density in units of nuclear matter
density, for the channels
(i) $D^+ D^-$, (ii) $D^0 \bar {D^0}$ as well as 
(iii) the total of these two channels.
For $eB=4 m_\pi^2$ and $eB= 8 m_\pi^2$, 
these are plotted for the symmetric nuclear matter ($\eta$=0)
and asymmetric nuclear matter (with $\eta$=0.5),
which may be compared with the results for the zero magnetic field 
shown in (a) and (b).
For $\eta=0$ (symmetric nuclear matter) and $eB=4 m_\pi^2$,
the decays $\psi (3686)$ to $D^+D^-$
and $\psi (3686)$ to $D^0 \bar {D^0}$ are observed to be possible
for densities above 4.3$\rho_0$ and 2.5$\rho_0$ respectively,
as can be seen in panel (c) of figure \ref{dwFT_psi3686_mag}. 
The masses of $D^+$ and $D^-$  mesons
have positive shifts due to the contributions from the 
lowest Landau levels \cite{dmeson_mag} in the presence of
magnetic field. This leads to the threshold 
density for the decay to $D^+D^-$ to be larger
as compared to the decay to $D^0 \bar {D^0}$
in symmetric nuclear matter ($\eta$=0).
For the larger value of the magnetic field ($eB=8 m_\pi^2$)
and for $\eta$=0, as shown in panel (e),
the decay of $\psi (3686)$ to the neutral $D^0 \bar {D^0}$ 
mesons becomes possible at a density of 2.3$\rho_0$, 
whereas the decay to the charged $D^+ {D^-}$ is not observed
even upto a density of 6$\rho_0$.
The in-medium decay widths in asymmetric nuclear
matter (with $\eta$=0.5) are plotted in the panels 
(d) and (f) for magnetic fields, $eB=4 m_\pi^2$
and  $eB=8 m_\pi^2$, which are compared to the case
of zero magnetic field, shown in (b). 
In the presence of asymmetry in the medium, 
the mass of the $D^0 \bar {D^0}$ pair is observed to be
larger as compared to the mass of the $D^+ D^-$ pair,
as can be seen from the masses of the $D$ and $\bar D$
mesons plotted in figures \ref{md_mag_eB} and \ref{mdbar_mag_eB}.
This leads to the production of the neutral open charm mesons
to be suppressed as compared to the $D^+D^-$ pair.
The threshold density above which the production of $D^0{\bar D^0}$
from decay of the charmonium state $\psi(3686)$ 
becomes possible ($\sim 3 \rho_0$) is observed to be larger 
than the production of $D^+D^-$ for isospin asymmetric nuclear matter,
in the absence of a magnetic field, as can be seen from
panel (b) of the figure \ref{dwFT_psi3686_mag}.
In the presence of a magnetic field, as
can be seen from the panels (d) and (f) of the
same figures, the threshold density is shifted 
to higher values for the production of $D^+D^-$ and
the decay to the neutral charm meson pair
is not kinematically possible. 

The in-medium decay widths of $\psi (3770)$ to $D\bar D$
are plotted in figure \ref{dwFT_psi3770_mag} as functions
of the baryon density for given values of the magnetic 
field for the symmetric as well as asymmetric nuclear matter. 
As has already been mentioned the decay of $\psi (3770)$ 
to $D\bar D$ takes place
in vacuum with the values of the decay widths
to $D^+D^-$ and $D^0 \bar {D^0}$ to be around
12 MeV and 16 MeV respectively.
At zero magnetic field \cite{amarvepja} and 
for symmetric nuclear matter,
there is observed to be an increase in the decay width
in both the channels $\psi(3770)\rightarrow D^+ D^-$
as well as $\psi (3770)$ to $D^0 \bar {D^0}$,
reaching a maximum value at around 0.5$\rho_0$,
followed by a drop upto around 2$\rho_0$, and 
a slow increase with further increase in density.
In asymmetric nuclear matter, the decay width
in the charged $D\bar D$ channel shows a
monotonic increase whereas the decay width to 
neutral $D\bar D$ mesons shows a drop
and then vanishing above densities of around 2.5 $\rho_0$.
the difference in the behaviours of the decay widths
of the charged and neutral $D\bar D$ channels 
should have modify the production of the 
open charm mesons in the compressed baryonic matter
experiments at the future facility at GSI, which should
have enhanced production of the $D^+D^-$ as compared
to the neutral $D\bar D$ mesons.

In the presence of magnetic field, the 
masses of the charged mesons ($D^+$ and $D^-$) 
have a positive shift due to the lowest Landau level
contribution and in symmetric nuclear matter,
the decay of $\psi(3770)$ to $D^+D^-$
is not observed upto a density (in units of $\rho_0$) 
of around 4.4 for $eB=4 m_\pi^2$,
as can be seen in (c) of figure \ref{dwFT_psi3770_mag}.
The decay of $\psi (3770)$ to $D^0 \bar {D^0}$ is
observed to decrease initially when the density 
is raised, upto a density of around 1.2 $\rho_0$,
followed by a rise with further increase in the
density. The behaviour remains the same for the 
case of the higher magnetic field, for 
the decay to $D^0 \bar {D^0}$, as can be seen
in panel (e) of figure \ref{dwFT_psi3770_mag}.
The decay to $D^+D^-$ is not observed for 
$eB=8 m_\pi^2$ for the symmetric nuclear matter
even upto a density of 6$\rho_0$. 

The behaviour of the decay widths of $\psi(3686)$ 
to $D^+D^-$ ($D^0 \bar {D^0}$) are very similar
to those calculated within the $^3P_0$ model
\cite{charm_decay_mag_3p0} as shown in figure
\ref{psi3686_decay_3p0_mag}, but the values for the
decay widths within the present model for composite hadrons,
are observed to be much smaller as compared 
to the results obtained using the $^3P_0$ model
\cite{charm_decay_mag_3p0}.

The decay widths of $\psi(3770)$ 
to $D^+D^-$ ($D^0 \bar {D^0}$) as calculated 
in the present work are compared with the 
results obtained using the the $^3P_0$ model
\cite{charm_decay_mag_3p0}, as shown in figure
\ref{psi3770_decay_3p0_mag}. The behaviour 
of these decay widths with density are observed
to be similar in both the models, however,
the values of the present model are observed to be
larger than those in the $^3P_0$ model.

The effect of isospin asymmetry is observed to be
important for both the cases of zero as well as
finite magnetic fields, and the production of
the $D^+D^-$ ($D^0 {\bar {D^0}})$ are observed
to be enhanced (suppressed) in the symmetric (asymmetric)
nuclear matter. This should have consequences on
the $D^+D^-$ production to be abundant as compared
to $D^0 \bar {D^0}$ from the asymmetric heavy ion
collisions in compressed baryonic matter (CBM)
experiments at FAIR, GSI.
At very small densities, the decay width 
of $\psi(3770)$ to $D^+D^-$ is observed to have 
a sharp increase (decrease) with density from 
the vacuum value for asymmetric nuclear matter 
in the absence (presence) of a magnetic field. 
Also, in the presence of a magnetic field, the
decay of $\psi (3770)$ to $D^0 \bar {D^0}$ 
is not kinematically possible for small densities 
in the asymmetric matter. These might have consequences 
on the production of the open charm mesons
from the charmonium state $\psi(3770)$ 
at the asymmetric heavy ion collision experiments 
at RHIC. 

\section{Summary}
The effects of magnetic field, isospin asymmetry
and baryon density on the partial decay widths
of the charmonium states ($J/\psi$, $\psi (3686)$,
$\psi (3770)$) to $D\bar D $ are studied
using a field theoretical model for composite hadrons
with quark (antiquark) constituents. The 
matrix element for the calculation of
the decay width is calculated from the free Dirac
Hamiltonian of the constituent quarks, using
the explicit constructions of the charmonium 
state, the $D$ and the ${\bar D}$ mesons.
The in-medium charmonium partial decay widths 
are computed from the medium modifications
of the charmonium and the open charm mesons
calculated using a chiral effective model.
The results of the present work are 
observed to have similar behaviour to the results
obtained using the $^3P_0$ model, but the values
of the decay widths obtained for decay of $\psi(3686)$
are seen to be much smaller in magnitude than those obtained
in the $^3P_0$ model. 
The charmonium decay to $D^+D^-$ and $D^0 \bar {D^0}$
are observed to be very different in the presence of 
magnetic field even in symmetric nuclear matter, 
due to the masses of the
charged mesons ($D^+$,$D^-$) which have a positive 
shift due to the Landau levels, whereas there 
is no such contribution to the masses of the
neutral $D^0$ and $\bar {D^0}$ mesons.   
The effects of isospin asymmetry give rise to
different modifications of the masses of the mesons
within the $D$ and $\bar D$ doublets.
The striking difference in the charmonium 
partial decay widths for the charged and neutral
$D\bar D$ pair channels in the isospin symmetric 
(asymmetric) nuclear matter will have 
consequences in the asymmetric heavy ion collisions
in the CBM experiments at FAIR, GSI as
a enhancement of the charged
$D^\pm$ mesons as compared to the neutral
$D\bar D$ mesons. At subnuclear densities,
there is observed to be a sharp drop for the
decay width of $\psi(3770)$ to $D^+D^-$
from its vacuum value in the presence 
of a magnetic field for asymmetric nuclear matter, 
and the decay to neutral $D\bar D$ is not 
kinematically possible.
These might have consequences on the production
of the charmonium state, $\psi(3770)$ as well as
open charm mesons in asymmetric heavy ion collisions
at RHIC. 

One of the authors (AM) is grateful to ITP, University of Frankfurt,
for warm hospitality and 
acknowledges financial support from Alexander von Humboldt Stiftung 
when this work was initiated. 


\vskip 0.15in
\noindent{\bf {Appendix A: Model for composite hadrons}}

\setcounter{equation}{0}
\renewcommand{\theequation}{A.\arabic{equation}}
\vskip 0.1in

The model describes hadrons comprising of 
quark and/or antiquark constituents.
The field operator for a constituent quark for a hadron at rest
at time, t=0, is written as
\begin {eqnarray}
\psi ({\bf x},t=0)
&=&(2\pi)^{-{3}/{2}}\int \Big [U({\bf k}) u_r q_r ({\bf k})
\exp(i{\bf k} \cdot{\bf x})
+ V({\bf k}) v_s \tilde q_s ({\bf k})
\exp(-i{\bf k} \cdot{\bf x})\Big ] d{\bfs k}\nonumber \\
 & \equiv & Q({\bf x})+\tilde Q({\bf x}),
\label{qx}
\end{eqnarray}
where, $U({\bfs k})$ and $V({\bfs k})$ are given as
\begin{eqnarray}
U({\bfs k})=\left (\begin{array}{c} f(|{{\bf k}}|)\\
{\bfm\sigma}\cdot {\bf k} g(|{{\bf k}}|)\\
\end{array} \right ),\;\;\;
V({\bfs k})=\left (\begin{array}{c} 
{\bfm\sigma}\cdot {\bf k} g(|{{\bf k}}|)\\
f(|{{\bf k}}|)\\
\end{array} \right ),
\label{ukvk}
\end{eqnarray}
The functions $f(|{\bf k}|)$ and $g(|{\bf k}|)$ satisfy the constraint
\cite{spm781},
$f^2+g^2 {\bf k}^2=1$,
as obtained from the equal time anticommutation relation 
for the four-component Dirac field operators. 
These functions, for the case of free Dirac field
of mass $M$, are given as,
\begin{equation}
f(|{\bf k}|)=\left ( \frac{k_0 +M}{2 k_0}\right )^{1/2},\;\;\;\; 
g(|{\bf k}|)=\left ( \frac{1}{2 k_0 (k_0+M)}\right )^{1/2},
\label{fkgk}
\end{equation}
where $k_0=(|{\bf k}|^2+M^2)^{1/2}$. In the above, $M$ is the constituent
quark/antiquark mass.
In equation (\ref{qx}), 
$u_r$ and $v_s$ are the two component spinors for the 
quark and antiquark respectively, satisfying the relations
$u_r^\dagger u_s=v_r^\dagger v_s=\delta_{rs}$.
The operator $q_{r}({\bf k})$ annihilates a quark with spin $r$ 
and momentum ${\bf k}$, whereas, $\tilde q _{s}({\bf k})$
creates an antiquark with spin $s$ and momentum ${\bf k}$,
and these operators satisfy the usual anticommutation relations
\begin{equation}
\{q_{r}({\bf k}),q_{s}({\bf k}')^\dagger\}=
\{\tilde q_{r}({\bf k}),\tilde q_{s}({\bf k}')^\dagger\}=
\delta _{rs} \delta ({\bf k}-{\bf k}').
\end{equation}

The field operator for the constituent quark of hadron with finite 
momentum is obtained by Lorentz boosting the field 
operator of the constituent quark of hadron at rest, which 
requires the time dependence of the quark field operators.
Similar to the MIT bag model \cite{MIT_bag}, 
where the quarks (antiquarks) occupy
specific energy levels inside the hadron, it is assumed 
in the present model for the composite hadrons that 
the quark/antiquark constituents carry  fractions
of the mass (energy) of the hadron at rest (in motion) 
\cite{spm781,spm782}.
The time dependence for the $i$-th quark/antiquark of a hadron
of mass $m_H$ at rest is given as
\begin{eqnarray}
Q_i(x)=Q_i({\bf x})e^{-i\lambda_i m_H t},\;\;
{\tilde Q}_i(x)={\tilde Q}_i({\bf x})e^{i\lambda_i m_H t},
\label{thadrest}
\end{eqnarray}
where $\lambda_i$ is the fraction of the energy (mass) of the hadron 
carried by the quark (antiquark), with $\sum_i \lambda_i=1$.
For a hadron in motion with four momentum p, 
the field operators for quark annihilation and antiquark creation,
for t=0, are obtained by Lorentz boosting the field operator of the 
hadron at rest, and are given as \cite{spmdiffscat}  
\begin{eqnarray}
Q^{(p)}({\bf x},t) =
\int \frac{d\bfs k}{(2\pi)^{{3}/{2}}}  
S(L(p)) U({\bf k}) 
Q ({\bf k}+\lambda {\bf p}) 
 \exp[{i({\bf k}+\lambda {\bf p})
\cdot{\bf x}-i\lambda p^0 t} ]
\label{qxp}
\end{eqnarray}
and,
\begin{eqnarray}
\tilde Q^{(p)}({\bf x},t) =
\int \frac{d\bfs k}{(2\pi)^{{3}/{2}}} 
S(L(p)) V(-{\bf k}) 
\tilde Q (-{\bf k}+\lambda {\bf p})
\exp[{-i(-{\bf k}+\lambda {\bf p}) \cdot{\bf x}
+i\lambda p^0 t}]. 
\label{tldqxp}
\end{eqnarray}
In the above, $\lambda$ is the fraction of the energy of the hadron, 
carried by the constituent quark (antiquark). 
In equations (\ref{qxp}) and (\ref{tldqxp}),
$L(p)$ is the Lorentz transformation matrix, which yields 
the hadron at finite four-momentum $p$ from the hadron at rest, 
and is given as \cite{spm782}
\begin{equation}
L_{\mu 0}=L_{0 \mu}=\frac {p^\mu}{m_H};\;\;\;\;\;
L_{ij}=\delta_{ij} +\frac {p^i p^j}{m_H (p^0+m_H)},
\label{lp}
\end{equation} 
where, $\mu=0,1,2,3$ and $i=1,2,3$,
and the Lorentz boosting factor $S(L(p))$ is given as 
\begin{equation}
S(L(p))=\Bigg [\frac {(p^0+m_H)}{2m_H}\Bigg ]^{1/2}
+\Bigg [ \frac {1}{2 m_H (p^0+m_H)} \Bigg ]^{1/2} 
{\vec {\alpha}}\cdot {\vec p},
\label{slp}
\end{equation}
where, $\vec \alpha = \left (
\begin{array}{cc}  0 & \vec \sigma \\ \vec \sigma & 0
\end{array}\right)$, are the Dirac matrices.
The Lorentz transformations used to obtain the constitutent quark and 
antiquark operators for hadron at rest to hadron
with momemtum, $p$, as given by equations (\ref{qxp}) 
and (\ref{tldqxp}) have the effect of addition of
the hadron fractional momentum, $\lambda {\bf p}$,
as a translation to the constituent quark (antiquark) 
momentum, ${\bf k}(-{\bf k})$ \cite{spmdiffscat}. 
This is similar to the quasipotential approach, 
where the Lorentz transformation plays the role of a 
translation \cite{quasi_pot_approach}.
Using the composite model picture with 
Lorentz transformations as considered 
in the present work, the various properties of 
hadrons, e.g., charge radii of the proton and pion, 
the nucleon magnetic moments \cite{spm781,spm782}, and, the
diffraction slopes as well as total cattering cross-sections
for the baryon-baryon and meson-baryon scattering have
been studied which have reasonable agreement with experiments
\cite{spmdiffscat}.  

The pair creation term of the Dirac Hamiltonian density 
\begin{equation}
{\cal H}_{Q^\dagger {\tilde Q}}(x)
=Q(x)^\dagger (-i{\bf {\alpha}}\cdot
{\bf \bigtriangledown} +\beta M)
{\tilde Q}(x) 
\label{pari_creation}
\end{equation}
is used to describe the strong decay of the hadron, $A$
at rest to $B({\bf p})$ and $C({\bf p'})$.
The operators for the quark and antiquark creation in the above term,
thus belong to different hadrons, $B$ and $C$ with 
4-momenta $p$ and $p'$ respectively. 
The light quark pair creation term of the Hamiltonian density,
is used to describe the decay of a heavy charmonium state
($\bar c c$) to $D$ and $\bar D$ states, which are
bound states of $c\bar q$ and $\bar c q$ repsectively,
with light ($u$, $d$) quark antiquark pair creation.
The present description of the decay of
$\Psi \rightarrow D \bar D$ using the pair creation
term of the Hamiltonian, as well as using an interaction term 
${\cal H}_I \sim \int d {\bf x} \bar \psi ({\bf x}) \psi ({\bf x})$
for the light quark pair creation term used in the $^3P_0$ model
\cite{3p0}, are consistent with the Okubo-Zweig-Iizuka (OZI) rule 
\cite{Okubo,Zweig,Iizuka}.

The composite model of hadrons with quark (antiquark)
constitutents, using the pair creation term of the
free Dirac Hamiltonian, as used in the present work, 
has been applied to study the strong decay processes
$\rho \rightarrow 2\pi$, $\Delta \rightarrow N\pi$,
$\phi \rightarrow K {\bar K})$,
$K^* \rightarrow K \pi$, which have good 
agreement with the experimental values of these decay widths
\cite{spm782}.

To compute the decay width of the charmonium state, $\Psi$
to $D\bar D$, we evaluate the matrix element of 
the quark-antiquark pair creation part of the Hamiltonian,
between the initial charmonium state and the final state for the reaction
$\Psi \rightarrow D ({\bf p})+{\bar D}({\bf p}')$.
The relevant part of the quark pair creation term is through the
$d \bar d (u \bar u)$ creation for decay to the final
state $D^+ D^-$ ($D^0 \bar {D^0}$). 
Using this pair creation term and the explicit expressions for
the states $|\Psi_m({\bf 0})\rangle$, $|D({\bf p})\rangle$
and $|{\bar D}({\bf p'})\rangle$ in terms of the quark and
antiquark constituents, we can evaluate 
\begin{eqnarray}
\langle D ({\bf p}) | \langle {\bar D} ({\bf p}')|
{\int {{\cal H}_{d^\dagger\tilde d}({\bf x},t=0)d{\bf x}}}
|{\Psi }_m (\vec 0) \rangle 
= \delta({\bf p}+{\bf p}')A^\Psi (|{\bf p}|)p_m.
\label{tfi}
\end{eqnarray}
With $\langle f | S |i\rangle =\delta_4 (P_f-P_i) M_{fi}$,
we have
\begin{equation}
M_{fi}=2\pi (-i A^ \Psi (|{\bf p}|)p_m.
\end{equation}
For evaluation of the  matrix element of 
the quark-antiquark pair creation part of the Hamiltonian,
between the initial charmonium state and the final state 
$D\bar D$ state as given by equation (\ref{tfi}),
As the $D$ and $\bar D$ mesons are nonrelativistic, 
we shall assume S(L(p)) and  S(L(p')) to be unity.
We shall also take the approximate forms (with a small momentum
expansion) for the functions $f(|{\bf k}|)$ and $g(|{\bf k}|)$ 
of the field operator as given by 
$g(|{\bf k}|)=1/\left ({2 k_0 (k_0+M)}\right )^{1/2}
\simeq {1}/({2M}),$
and $f(|{\bf k|})=(1-g^2{\bf k}^2)^{1/2} 
\approx 1-((g^2 {\bf k}^2)/2)$ \cite{amspmwg}. 

The expression for the decay width  is obtained as
\begin{eqnarray}
&&\Gamma(\Psi\rightarrow D({\bf p}) {\bar D} (-{\bf p}))
 \nonumber \\
&=&\gamma_\Psi^2 \frac{1}{2\pi} 
\int \delta(m_{\Psi}-p^0_{D}-p^0_{\bar D})
|M_{fi}|^2_{\rm {av}}
\cdot 4\pi |{\bfs p}_{D}|^2 d|{\bfs p}_{D}| 
\nonumber\\
&=& \gamma_\Psi^2\frac{8\pi^2}{3}{\bf p}|^3
\frac {p^0_{D} p^0_{\bar D}}{m_{\Psi}}
A^{\Psi}(|{\bf p}|)^2
\label{gammapsiddbar}
\end{eqnarray}
In the above, $p^0_{D ({\bar D})}
=\big(m_{D ({\bar D})}^2+{\bf p}^2\big)^{\frac{1}{2}}$, 
and, $|\bfs p|$ is the magnitude of the momentum of the outgoing 
$D ({\bar D})$ mesons. 
The expression for $A^\Psi(|\bf p|)$ 
in the above equation is given as
\begin{eqnarray}
A^{\Psi}(|{\bf p}|) = 6c_\Psi\exp[(a_\Psi {b_\Psi}^2
-R_D^2\lambda_2^2){\bf p}^2]
\cdot\Big(\frac{\pi}{a_\Psi}\Big)^{{3}/{2}}
\Big[F_0^\Psi+F_1^\Psi\frac{3}{2a_\Psi}
+F_2^\Psi\frac{15}{4a_\Psi^2}\Big],
\label{ap}
\end{eqnarray}
where $a_\Psi$, $b_\Psi$
are given as \cite{amspmwg}
\begin{equation}
a_\Psi=\frac{1}{2}R_{\Psi}^2+R_D^2; \;\;\;\; 
b_\Psi=R_D^2\lambda_2/a_\Psi,
\label{abpsi}
\end{equation}
and $c_\Psi$, $\Psi=J/\psi,\psi',\psi"$, are given
by the expressions
\begin{equation}
c_{J/\psi}=\frac{1}{\sqrt{6}}\cdot
\left(\frac{R_\psi^2}{\pi}\right)^{{3}/{4}}
\cdot\frac{1}{6}\cdot\left(\frac{R_D^2}{\pi}\right)
^{{3}/{2}},
\label{cpsi}
\end{equation}
\begin{equation}
c_{\psi'}=\frac{1}{\sqrt{6}}\left(\frac{3}{2}\right)^{{1}/{2}}
\left(\frac{R_{\psi'}^2}{\pi}\right)^{{3}/{4}}
\cdot\frac{1}{6}\cdot\left(\frac{R_D^2}{\pi}\right)^{{3}/{2}},
\label{cpsip}
\end{equation}
\begin{equation}
c_{\psi''}=\frac{1}{4\sqrt{3\pi}}
\left ({\frac{16}{15}}\right)^{{1}/{2}}
\cdot \pi^{-{{1}/{4}}}
\cdot (R_{\psi''}^2)^
{{7}/{4}}\cdot\frac{1}{6}\cdot
\left(\frac{R_D^2}{\pi}\right)^{{3}/{2}}.
\label{cpsipp}
\end{equation}
In the above expressions, $R_\Psi$ and $R_D$ refer to the 
strengths of the harmonic oscillator wave functions for the
charmonium state, $\Psi (J/\psi, \psi(3686), \psi(3770))$ 
and the $D(\bar D)$ mesons,
whose explicit constructions are given by equations
(\ref{Psi}), (\ref{d}) and (\ref{dbar}).
The expressions for 
$F_0^\Psi$,  $F_1^\Psi$,  $F_2^\Psi$, 
for $\Psi\equiv J/\psi,\psi',\psi''$, are given as
\begin{eqnarray}
&&F^{J/\psi}_0=(\lambda_2-1)-2g^2\bfs p^2(b_{J/\psi}-\lambda_2)
\nonumber \\
&\times&
\left(\frac{3}{4}b_{J/\psi}^2-(1+\frac{1}{2}\lambda_2)b_{J/\psi}
+\lambda_2-\frac{1}{4}\lambda_2^2\right),
\nonumber\\
&&F^{J/\psi}_1=g^2\left[-\frac{5}{2}b_{J/\psi}+\frac{2}{3}
+\frac{11}{6}\lambda_2\right],
\nonumber\\
&&F^{J/\psi}_2=0,
\label{c012psi}
\end{eqnarray}
\begin{eqnarray}
F^{\psi'}_0&=&\left(\frac{2}{3}R_{\psi'}^2{b_{\psi'}}^2\bfs p^2-1\right)
F^{J/\psi}_0,
\nonumber\\
F^{\psi'}_1&=&\frac{2}{3}R_{\psi'}^2 F^{J/\psi} _0
+\left(\frac{2}{3}R_{\psi'}^2 b_{\psi'}^2\bfs p^2
-1\right)
F^{J/\psi}_1
\nonumber \\ &-&\frac{8}{9}R_{\psi'}^2{b_{\psi'}}g^2\bfs 
p^2\left[\frac{9}{4}b_{\psi'}^2-b_{\psi'}\left(2+\frac{5}{2}\lambda_2
\right)+2\lambda_2+\frac{1}{4}\lambda_2^2\right],\nonumber\\
F^{\psi'}_2&=&\frac{2}{3}R_{\psi'}^2g^2
\left[-\frac{7}{2}b_{\psi'}+\frac{2}{3}+\frac{11}{6}
\lambda_2\right],
\label{c012psip}
\end{eqnarray}
and,
\begin{eqnarray}
F_0^{\psi''}&=&2b_{\psi''}^2(1-\lambda_2)\bfs p^2
+2b_{\psi''}^2g^2(\bfs p^2)^2(b_{\psi''}-\lambda_2)((3/2)b_{\psi''}^2
\nonumber \\
&-&(2+\lambda_2)b_{\psi''}+2\lambda_2-(1/2)\lambda_2^2),\nonumber\\
F_1^{\psi''}&=&g^2\bfs p^2[14{b_{\psi''}}^3-b_{\psi''}^2((32/3)
+(37/3)\lambda_2)\nonumber \\
&+&b_{\psi''}((28/3)\lambda_2-(1/3)\lambda_2^2)],
\nonumber\\
F_2^{\psi''}&=&g^2[7b_{\psi''}-(2/3)\lambda_2-(4/3)].
\label{c012psipp}
\end{eqnarray}

In the expressions for the decay widths of the charmonium state,
$\Psi$ decaying to $D^+D^- (D^0{\bar {D^0}})$, 
the parameter, $\gamma_\Psi$ is introduced, which refers to
the production strength of $D\bar D$ from decay 
of charmonium $\Psi$ through light quark 
pair creation. In the $^3P_0$ model, such a
light quark-antiquark pair creation strength parameter, 
$\gamma$ has been introduced  \cite{amarvepja,3p0,friman}. 
The parameter, $\gamma$ in the $^3P_0$ model,
as well as the parameter $\gamma_\Psi$ in the model
for composite hadrons \cite{amspmwg} used in the present work,
are chosen so as to reproduce the vacuum decay widths for 
the decay channels $\psi'' \rightarrow D^+ D^-$ and 
$\psi'' \rightarrow D^0 \bar {D^0}$.

\vskip 0.2in
\noindent{\bf {Appendix B: Charm mesons in the chiral effective model}}

\setcounter{equation}{0}
\renewcommand{\theequation}{B.\arabic{equation}}

We study the medium modifications of the masses
of the open charm ($D$ and $\bar D$) mesons and the charmonium
states in the presence of a magnetic field,
within a chiral effective model
\cite{amarvepja,dmeson_mag,charmonium_mag}.
The model is a generalization of a chiral SU(3) model
to SU(4) so as to include the interactions of the charmed
mesons with the light hadronic sector. 

The Lagrangian density of the chiral SU(3) model, in the presence
of magnetic field, is given as \cite{paper3,kmeson_mag}
\bea
{\cal L} = {\cal L}_{kin} + \sum_ W {\cal L}_{BW}
          +  {\cal L}_{vec} + {\cal L}_0 
+ {\cal L}_{scalebreak}+ {\cal L}_{SB}+{\cal L}_{mag}^{B\gamma},
\label{genlag} \eea
where, $ {\cal L}_{kin} $ corresponds to the kinetic energy terms
of the baryons and the mesons,
${\cal L}_{BW}$ contains the baryons with the meson,
$W$ (scalar, pesudoscalar, vector, axialvector meson),
$ {\cal L}_{vec} $ describes the dynamical mass
generation of the vector mesons via couplings to the scalar fields
and contains additionally quartic self-interactions of the vector
fields, ${\cal L}_0 $ contains the meson-meson interaction terms
${\cal L}_{scalebreak}$ is a scale invariance breaking logarithmic
potential and $ {\cal L}_{SB} $ describes the explicit chiral symmetry
breaking. The term ${\cal L}_{mag}^{B\gamma}$, describing the interacion
of the baryons with the electromagnetic field is given as 
\cite{dmeson_mag,kmeson_mag}
\be 
{\cal L}_{mag}^{B\gamma}=-{\bar {\psi_i}}q_i 
\gamma_\mu A^\mu \psi_i
-\frac {1}{4} \kappa_i \mu_N {\bar {\psi_i}} \sigma ^{\mu \nu}F_{\mu \nu}
\psi_i,
\label{lmag_Bgamma}
\ee
where, $\psi_i$ corresponds to the $i$-th baryon, with electric charge
$q_i$, $A^\mu$ is the electromagnetic field, $F^{\mu \nu}$
is the electromangetic field tensor and $\mu_N$ is the nuclear magneton.
The tensorial interaction of baryons 
with the electromagnetic field given by the second term 
of the above equation accounts for the anomalous magnetic moments 
of the baryons (through the values of $\kappa_i$).
The charged baryons have contributions from the Landau 
energy levels.

The concept of broken scale invariance of QCD
is simulated in the effective 
Lagrangian model through the introduction of 
the scale breaking term \cite{paper3}
\begin{equation}
{\cal L}_{scalebreak} =  -\frac{1}{4} \chi^{4} {\rm {ln}}
\Bigg ( \frac{\chi^{4}} {\chi_{0}^{4}} \Bigg ) + \frac{d}{3}{\chi ^4} 
{\rm {ln}} \Bigg ( \bigg (\frac{I_{3}}{{\rm {det}}\langle X 
\rangle _0} \bigg ) \Bigg ),
\label{scalebreak}
\end{equation}
where $I_3={\rm {det}}\langle X \rangle$, with $X$ as the multiplet
for the scalar mesons. 
Equating the trace of the energy momentum tensor of QCD
in the massless quarks limit to that of the chiral 
effective model, and, using the one loop beta function
for $N_c$=3 and $N_f$=3, leads to the relation of the 
scalar gluon condensate to the dilaton field 
as given by 
\begin{equation}
\left\langle  \frac{\alpha_{s}}{\pi} G_{\mu\nu}^{a} G^{ \mu\nu a} 
\right\rangle =  \frac{8}{9}(1 - d) \chi^{4}.
\label{chiglu}
\end{equation}

The scalar fields (non-strange scalar-isoscalar, field $\sigma$, 
non-strange scalar isovector, $\delta$, the strange field
$\zeta$) and the scalar dilaton field, $\chi$ are solved 
from their equations of motion in isospin asymmetric
nuclear matter in the presence of magnetic fields.
For given values of the baryon density, $\rho_B$, the isospin 
asymmetry parameter, $\eta= ({\rho_n -\rho_p})/({2 \rho_B})$
(where $\rho_n$ and $\rho_p$ are the number densities of the neutron
and the proton respectively) and magnetic field, the 
values of the scalar fields are obtained. 
We neglect the small isospin effect arising from the
difference in the current quark masses of the u and d
quarks. The calculations for the scalar fields are carried out in
the mean field approximation, where the meson fields are replaced
by their expectation values. 
In addition, we also use the approximations that
$\bar \psi_i \psi_j = \delta_{ij} \langle \bar \psi_i \psi_i
\rangle \equiv \delta_{ij} \rho_i^s $ and 
$\bar \psi_i \gamma^\mu \psi_j = \delta_{ij} \delta^{\mu 0} 
\langle \bar \psi_i \gamma^ 0 \psi_i
\rangle \equiv \delta_{ij} \delta^{\mu 0} \rho_i $, 
where, $\rho_i^s$ and $\rho_i$ are the scalar 
and number density of $i$-th baryon. 
The values of the scalar fields in the (magnetized) hadronic
medium are used to obtain the in-medium masses of the $D$ and $\bar D$
mesons as well as the charmonium states
\cite{dmeson_mag,charmonium_mag,charm_decay_mag_3p0}.

The open charm $D$ and $\bar D$ mesons are studied
within the chiral model by generalizing chiral SU(3) 
to chiral SU(4) to obtain the interactions of the charmed 
mesons with the light hadrons. 
As the baryons belong to 20-plet and mesons to the 16-plet
representation in SU(4), the baryons are represented by
the tensor $B^{ijk}$ (which are antisymmetric in the first
two indices), to derive the interactions of the pesudoscalar
$D$ and $\bar D$ mesons with the baryons \cite{amarvepja}.
The in-medium masses of these open charm mesons 
are computed from their interactions with the baryons and 
the scalar mesons, arising from the 
leading Weinberg-Tomozawa vectorial interaction
with the baryons, the scalar exchange interactions and
the range terms \cite{amarvepja}. 
The dispersion relations for the $D$ and $\bar D$ mesons
are obtained from the Fourier transformations of the
equations of motion of these mesons.
In the presence of the magnetic field, the
charged $D^\pm$ mesons have positive contributions
to their masses from the Landau levels,
and we retain the lowest Landau level contribution
in the masses of the charged $D$ and $\bar D$ mesons.

The mass shift of the charmonium state, $\Psi$ 
($J/\psi$, $\psi(3686)$ and $\psi(3770)$) 
is obtained from the medium modification of the
scalar gluon condensate, which is computed from the
medium change of the dilaton field calculated 
within the chiral effective model, using the equation
(\ref{chiglu}). The shift in the charmonium mass is given as
\cite{amarvdmesonTprc,amarvepja}
\begin{equation}
\Delta m_{\Psi}= \frac{4}{81} (1 - d) \int dk^{2} 
\langle \vert \frac{\partial \psi (\vec k)}{\partial {\vec k}} 
\vert^{2} \rangle
\frac{k}{k^{2} / m_{c} + \epsilon}  \left( \chi^{4} - {\chi_0}^{4}\right), 
\label{masspsi}
\end{equation}
where 
\begin{equation}
\langle \vert \frac{\partial \psi (\vec k)}{\partial {\vec k}} 
\vert^{2} \rangle
=\frac {1}{4\pi}\int 
\vert \frac{\partial \psi (\vec k)}{\partial {\vec k}} \vert^{2}
d\Omega,
\end{equation}



\end{document}